\documentclass[runningheads]{llncs}
\usepackage{graphicx}
\usepackage{amsmath,amssymb,amsfonts,mathtools}

\newcommand{\RomanNumeralCaps}[1]
{\MakeUppercase{\romannumeral #1}}
\usepackage{algorithm}
\usepackage{algorithmic}
\usepackage{grffile}
\usepackage{cuted}
\usepackage{bm}
\usepackage{mathrsfs}
\usepackage{lipsum}
\usepackage{color}
\usepackage{lscape}
\usepackage{multirow} 

\usepackage{booktabs}

\usepackage{multirow}
\usepackage{threeparttable}
\usepackage{textcomp}

\usepackage[misc]{ifsym}
\usepackage{graphicx}
\newtheorem{observation}{Observation}
\begin{document}
\title{Online Ride-Hitching in UAV Travelling\thanks{A preliminary version of this paper is to appear in COCOON 2021.}}
%
%
\author{Songhua Li (\Letter)\inst{1}\and Minming Li\inst{1} \and  Lingjie Duan\inst{2} \and Victor C.S. Lee\inst{3}}
%

%
\institute{Department of Computer Science, City University of Hong Kong, Kowloon, Hong Kong SAR, China \\
\email{songhuali3-c@my.cityu.edu.hk, minming.li@cityu.edu.hk}
\and
Engineering Systems and Design Pillar, Singapore University of Technology and Design, Singapore
\\
\email{lingjie\_duan@sutd.edu.sg}
\and 
Department of Electrical and Electronic Engineering, The University of Hong Kong, Hong Kong SAR, China\\
\email{csvlee@eee.hku.hk}
}
\authorrunning{S. Li et al.}
%

%
\maketitle              
\begin{abstract}
The unmanned aerial vehicle (UAV) has emerged as a promising solution to provide delivery and other mobile services to customers rapidly, yet it drains its stored energy quickly when travelling on the way and (even if solar-powered) it takes time for charging power on the way before reaching the destination. 
To address this issue, existing works focus more on UAV's path planning with designated system vehicles providing charging service. However, in some emergency cases and rural areas where system vehicles are not available, public trucks can provide more feasible and cost-saving services and hence a silver lining. In this paper, we explore how a single UAV can save flying distance by exploiting public trucks, to minimize the travel time of the UAV. We give the first theoretical work studying online algorithms for the problem, which guarantees a worst-case performance. 
We first consider the offline problem knowing future truck trip information far ahead of time. By
delicately transforming the problem into a graph satisfying both time and power constraints, we present 
a shortest-path algorithm that outputs the optimal solution of the problem.
Then, we proceed to the online setting where trucks appear in real-time and only inform the UAV of their trip information some certain time $\Delta t$ beforehand. As a benchmark, we propose a well-constructed lower bound that an online algorithm could achieve.  We propose an online algorithm \textsc{MyopicHitching} that greedily takes truck trips and an improved algorithm \textsc{Adaptive} that further tolerates a waiting time in taking a ride. Our theoretical analysis shows that \textsc{Adaptive} is asymptotically optimal in the sense that its ratio approaches the proposed lower bounds as $\Delta t$ increases.
\keywords{Ride-hitching \and Energy efficiency \and Online algorithm.}
\end{abstract}
\section{Introduction}
 As technologies in navigation and control progress, the application of unmanned aerial vehicle (UAV) in package delivery is proved to be a promising approach. For example, the e-commerce giant Amazon has been pushing to deliver packages to its millions of customers by drones. DHL applies drones to provide fully autonomous loading and offloading in the last-mile delivery \cite{b33}, which provides a silver lining in special situations (e.g., the COVID-19 pandemic) with social-distancing. 
 However, due to the nature of the UAV/drone in both the low energy storage and high-rate flying consumption, it quickly drains its stored energy, which limits the delivery range and affects the service effectiveness remarkably. Although the UAV can be solar-powered by utilizing solar radiation as energy \cite{b32}, this is not sufficient since the charging-rate is not high enough.
 Fortunately, the UAV is able to dock with road vehicles automatically \cite{bokeno2018package}, which makes it possible for UAV to team up with trucks spontaneously and instantaneously for reducing the transportation cost. This is inspired by ride-sharing platforms, for example, GrabHitch allows passengers to hitch ride on the way. However, it is still not clear what ramifications of online ride-hitching are on UAV's energy saving. This paper aims to provide theoretical foundations for online ride hitching in UAV travelling. We note that the UAV may not catch a truck which is far away from its current location (\textit{spatial issue}), nor wait for a truck for too long (\textit{temporal issue}) as time efficiency is critical in UAV's delivery service. 

\textbf{Related works.} 
We survey relevant researches along two threads. \textit{The first thread} studies UAV's energy-efficiency problem with either routing or speed scheduling optimization. For example, \cite{b35} proposed the \textit{looking before crossing} algorithm, which is proved to be optimal for the offline speed scheduling problem under a practical flight energy model. \cite{b37} considers an energy-aware path planning algorithm that minimizes energy consumption while satisfying coverage and resolution constraints. Please refer to \cite{b34} for a survey work. In contrast, we focus on theoretical issues of the problem especially when truck trips are released in an online fashion. We aim to unveil the adaptability of the "ride-sharing" in UAV travelling in the worst-case scenario, which is usually measured by online algorithms and competitive ratio \cite{b10}. 
\textit{The second thread} focuses on classical combinatorial optimization problems. The $k$-server problem aims to efficiently move $k$ servers to serve a batch of online requests of the metric space \cite{{add2}} such that the total moving distance of all servers is minimized. The famous work function algorithm achieves a competitive ratio of $2k-1$ on general metrics and hence is optimal for $1$-server problem on the line \cite{add3}. When the server moves in constant velocity, the work function algorithm is optimal in achieving minimum completion time of serving all the online requests. A variant of the 1-server problem is the online repairman problem \cite{add1} which asks for a tour that visits a set of online cities in the metric space such that the weighted sum of completion times of the cities is minimized. \cite{add1} proposes a $(1+\sqrt{2})^2$-competitive deterministic online algorithm for the general metric spaces and \cite{add4} gave an improved 5.429-competitive algorithm for line metrics. A generalized version of the $k$-server problem is the $k$-taxi problem in which each request is represented by a pair $(s,t)$ of two points (including the start point $s$ and end point $t$) instead \cite{add5}. In the gas station problem \cite{gasstation}, a vehicle with a given tank capacity $U$ and an initial amount $\mu_s$ of gas, can purchase gas at each vertex of a complete graph at a certain price. And the objective of \cite{gasstation} is to find 
the cheapest way from a given start node $s$ to a given target node $t$ of the graph. Note that 
the gas station problem involves neither the time constraint on vertices nor restrictions in the set of visited vertices. Another related problem is the online maximum $k$-interval coverage problem which aims to select $k$ online sub-intervals (i.e., truck trips) such that the total covered length of a target interval (i.e., the UAV's path) is maximized \cite{add6}. 
In contrast, the problem studied in this paper is more complicated since one has to face both the \textit{spatial} and the \textit{temporal issues} simultaneously and the power constraint of UAV travelling is further involved.

Main contributions of this paper are summarized below. Due to space constraints, some results and proofs are deferred to the Appendix.
\begin{itemize}
    \item We are the first to study UAV's traveling problem by hitching on truck rides in an online setting, for the purpose of minimizing the UAV's travel time. Comprehensively, we investigate different cases according to how early (i.e., $\Delta t$) a truck should inform the UAV of its trip before the departure.
    \item  For the offline version of the problem, we give a graph-based optimal solution. Since it is intricate to capture both power and time constraints in mapping truck trips to nodes of a graph, we delicately construct the graph by screening unnecessary trucks iteratively, which is on top of some characteristics of the problem. Based on this, we find an optimal solution in $O(n^2)$ time.
    \item As a benchmark, we construct lower bounds on the competitive ratio for any online algorithms, by considering different time gap $\Delta t$ between the start time and release time.
    \item  We show that a simple myopic algorithm (where the UAV flies forward constantly by default until using up its stored energy, and myopically accepts as many rides as possible halfway) has a defect, which can be easily exploited by the adversary, leading to negligible energy saved from taking rides. To fix this defect, we propose a $\Delta t$-\textsc{Adaptive} algorithm by tolerating a waiting time at most $\frac{\Delta t}{2}$ in taking each ride, which achieves a provable competitive ratio very close to the lower bound. 
\end{itemize}
\section{Problem Formulation}
We consider the following problem: the UAV, which is at its origin $O$, is supposed to move to its destination $A$ as early as possible, in which the path length $|OA|=a$. The UAV has a low charging rate $\alpha$ per unit time and a high power-consuming rate $\beta$ ($>\alpha$) per unit time.  Initially at time $0$, the UAV stores an amount $P_0$ of energy (which is small and could be zero) and it flies at its maximum velocity
\footnote{In practice, the UAV can convert its speed between 0 and its maximum speed in few seconds \cite{b37}. Considering the dominant travel time, we assume the flying velocity as a constant as in \cite{b40,b37}.} $v_0$. To avoid the trivial case that the UAV directly flies to the destination by charging on the way, we assume the UAV has insufficient energy $P_0$ ($<\frac{(\beta-\alpha)a }{v_0}$). That is, the UAV needs to hitch truck rides to save flying distance or charge for sufficiently large amount of time to fly to the destination. Along the path $OA$ of the UAV, a sequence $\mathbb{V}=\{V_1, V_2,..., V_n\}$ of $n\in \mathbb{N}^+$ trucks will be released one by one to potentially offer rides to the UAV. Each truck $V_i=(r_i,t_i,o_i,d_i,v_i)\in \mathbb{V}$ releases its trip information to the UAV at time $r_i$ and departs from its origin $o_i$ at time $t_i$ ($\geq r_i$) with a constant velocity $v_i$ to its destination $d_i$. Further, we denote $\Delta t_i=t_i-r_i$ as the time gap between the start time $t_i$ and the release time $r_i$ of a truck ride $V_i$. The UAV is not informed of each ride $V_i\in\mathbb{V}$ until its release time $r_i$ (when the truck's schedule is determined) and needs to determine whether to accept/catch or reject $V_i$ irrevocably at $r_i$. The \textit{objective} is to minimize the UAV's travel time to the destination $A$ (or equivalent, the arrival time at $A$) by using online truck rides to save energy halfway. Key notations of this paper are given below in Table \ref{notatinosinthispaper}. 
\begin{table}
\caption{Notations in this paper.}\label{notatinosinthispaper}
\begin{center}
\begin{tabular}{cp{260pt}}
\hline \hline
 Notations & Physical meanings\\
 \hline \hline
 $[0,a]$& The line segment representing the UAV's route.\\
\hline
$\mathbb{V}_i=\{V_1,V_2,...,V_i\}$ & The sequence of the first $i$ rides released, particularly, $\mathbb{V}_n=\mathbb{V}$.\\
\hline
$V_i=(r_i,t_i,o_i,d_i,v_i)$&  The $i$th released ride, with its release time $r_i$, departure time $t_i$, start location $o_i$, end location $d_i$ and flying velocity $v_i$.\\
 \hline 
 $\Delta t_i=t_i-r_i$& the time gap between the start time $t_i$ and the release time $r_i$ of a ride $V_i$.\\
 \hline
 $v=\min\limits_{V_i\in\mathbb{V}}v_i$&the smallest possible velocity of a truck/ride.\\
\hline
$P_0$& The initial power that the UAV contains at time $t_0=0$.\\
\hline
$P_i$ &The power that the UAV contains at the start time of the $i$th ride taken by the UAV.\\
\hline
$\beta$&The UAV's power-consumption-rate for flying.\\
\hline 
$\alpha$&The UAV's recharging-rate, which satisfies $\alpha<\beta$;\\
\hline
$\xi(U)$ and $\xi(V_{|U|})$ & The arrival time (or the overall travel time) of the UAV to the target $A$, given the set $U$ of rides and the last ride $V_{|U|}$ taken by the UAV, respectively, as formally defined later in (\ref{arrivaltimelimit}) and (\ref{arrivaltimelimittwo}).\\
\hline \hline
\end{tabular}
\end{center}
\end{table}

Suppose that $U=(V_1,\cdots, V_{|U|})$ is the sequence of rides to be taken by the UAV. Denote $P_i$ as the power that the UAV contains at the start time $t_i$ of the $i$th taken ride $V_{i}$. Now, we formally formulate \textit{our model} as the following mathematical problem (\ref{modelobjlimit001})-(\ref{model2const13}). 
where the objective (\ref{modelobjlimit001}) is to minimize the UAV's travel time from $O$ to $A$, which is according to the following Proposition \ref{objinlimittransfer} where we discuss the physical meaning of the two terms of (\ref{modelobjlimit001}) and why we take the maximizing operation between the two. Constraints (\ref{model2const10})-(\ref{model2const11}) indicate the power $P_{i+1}$ that the UAV contains at the start time $t_{i+1}$ of $V_{i+1}$ by transferring from $V_i$, which can be calculated  by the following \textit{power transfer function} in (\ref{powertransferfunctionnew}).
\begin{equation}\label{powertransferfunctionnew}
\textsc{PTF}(P_i,V_i,V_{i+1})=P_i+(t_{i+1}-t_i)\cdot\alpha-\frac{|o_{i+1}-d_i|\beta}{v_0} 
\end{equation}
After leaving each truck $V_i$, for $i\in\{1,\cdots,|U|-1\}$, note that the UAV
needs to have enough energy to fly to the following truck $V_{i+1}$'s start location, which leads to the \textit{power compatibility} constraint (\ref{model2const12}) as $\textsc{PTF}(P_i,V_i,V_{i+1})\geq 0$; and it also needs to catch $V_{i+1}$'s start time, which is reflected by the \textit{time compatibility} constraints (\ref{model2const2})-(\ref{model2const7}). When the inequality in constraint (\ref{model2const7}) holds strictly, the UAV needs to stop at the roadside to wait for $V_{i+1}$'s  departure \footnote{We use stop-and-recharge time and waiting time interchangeably in this paper, to refer to the time that the UAV stops at roadside.}  after leaving $V_i$.  Constraint (\ref{model2const13}) indicates the total flying distance of the UAV.
\begin{alignat}{2}
    \min\limits_{U\subseteq \mathbb{V}}\max\{& \lambda \frac{\beta}{\alpha v_0}-\frac{P_0}{\alpha},t_{|U|}+\frac{d_{|U|}-o_{|U|}}{v_{|U|}}+\frac{a-d_{|U|}}{v_0}\}&\label{modelobjlimit001}\\
\mbox{s.t.}\quad
  & P_1=P_0+t_1\cdot \alpha-\frac{o_1}{v_0}\beta
   &{}&\label{model2const10}\\
&P_{i+1}= \textsc{PTF}(P_i,V_i,V_{i+1})
  & {}&\label{model2const11}\\
 &\textsc{PTF}(P_i,V_i,V_{i+1})\geq 0& {}&\label{model2const12}\\
  &t_1\geq \frac{o_1}{v_0}&{}&\label{model2const2}\\
  &t_{i+1}\geq t_i+\frac{d_i-o_i}{v_i}+\frac{|o_{i+1}-d_i|}{v_0}&  {}  &\label{model2const7}\\
 &\lambda =(o_1+\sum_{i=1}^{|U|-1} |o_{i+1}-d_i|+a-d_{|U|})&{}&\label{model2const13}
 \end{alignat}
For analytical tractability, the overall travel time of the UAV in  objective (\ref{modelobjlimit001}) is converted to  (\ref{arrivaltimelimit}) in the following Proposition  \ref{objinlimittransfer}. Intuitively, when the UAV contains enough power to fly constantly to $A$ after completing the last-taken ride $V_{|U|}$, UAV's arrival time only corresponds to $V_{|U|}$; otherwise, UAV's arrival time to $A$ only corresponds to the UAV's overall flying distance. Thus, we have
\begin{proposition}\label{objinlimittransfer}
Given the sequence $U=(V_1,\cdots,V_{|U|})$ of rides taken by the UAV, the UAV's arrival time $\xi(U)$ at the target $A$ is given by
\begin{small}
\begin{equation}\label{arrivaltimelimit}
\xi(U)=\max\{\underbrace{(o_1+\sum_{i=1}^{|U|-1} |o_{i+1}-d_i|+a-d_{|U|})}_{\rm i.e.,\; UAV's\;overall\;flying\;distance}\frac{\beta}{\alpha v_0}-\frac{P_0}{\alpha},
t_{|U|}+\frac{d_{|U|}-o_{|U|}}{v_{|U|}}+\frac{a-d_{|U|}}{v_0}\}
\end{equation}
\end{small}
\end{proposition}

\textbf{Competitive Ratio}. Online algorithms are typically measured by the competitive ratio \cite{b10}. Given a sequence $\mathbb{V}$ of online trucks that can offer rides to the UAV, denote by $\xi_{\rm ALG}(\mathbb{V})$ and $\xi_{\rm OPT}(\mathbb{V})$ the UAV's arrival time to the destination by an online algorithm (ALG) and the optimal offline solution (OPT) where complete information of $\mathbb{V}$ is given beforehand, respectively. Then, the competitive ratio $\rho$ of the problem is defined as $\rho=\max\limits_{\mathbb{V}}\frac{\xi_{\rm ALG}(\mathbb{V})}{\xi_{\rm OPT}(\mathbb{V})}$. When a number $\theta\geq 1$ satisfies $\theta\leq \rho$ for all deterministic online algorithms, we say $\theta$ is a lower bound on the competitive ratio of the problem. 
\section{Offline Problem And Algorithm Design}\label{limittransfermodels}
We present an optimal solution, named \textsc{OptimalHitching}, for the offline problem in this section, in which the idea behind is to map the offline rides to nodes in a graph, and further the taking sequence in achieving the earliest arrival time is converted to a minimum-weight path in the constructed graph. In a graph-based solution, we find it is difficult to map the UAV's arrival time to edge weight in the graph directly by using $\xi(U)$. This is because the arrival time $\xi(U)$ corresponds to multi nodes/rides in the set $U$. Hence, we transform the objective  (\ref{arrivaltimelimit}) to the following (\ref{arrivaltimelimittwo}), which only corresponds to the last ride $V_{|U|}$ taken by the UAV and the power $P_{|U|}$ that the UAV contains at the start time $t_{|U|}$ of $V_{|U|}$.
\begin{small}
\begin{equation}\label{arrivaltimelimittwo}
\xi(V_{|U|})= t_{|U|}+\frac{d_{|U|}-o_{|U|}}{v_{|U|}}+\frac{a-d_{|U|}}{v_0}+\underbrace{\frac{\max\{\frac{a-d_{|U|}}{v_0}(\beta-\alpha)-(P_{|U|}+\frac{d_{|U|}-o_{|U|}}{v_{|U|}}),0\}}{\alpha}}_{\rm the\; possible\;stop-and-recharge\;time\;of\;the\;UAV}
\end{equation}
\end{small}
Before going into the details of our offline algorithm, we first give the following definitions together with some preliminary results. 
\begin{definition}[Adjacent]
Given the set $U\subset\mathbb{V}$ of rides accepted by the UAV, two trips $V_i$ and $V_j$ in $U$ are called adjacent rides if and only if
$V_j\in\{\arg\min\limits_{\mathclap{V_x\in U, o_x>o_i}}\{o_x-o_i\},\;\arg\max\limits_{\mathclap{V_x\in U, o_x<o_i}}\{o_x-o_i\}\}$.
    Specifically, $V_j$ is regarded as the prior one (resp. the following one) of the two if $o_j<o_i$ (resp. $o_j>o_i$).
\end{definition}
\begin{definition}[Sequentially-taken]
Given the set $U\subset\mathbb{V}$ of rides taken by the UAV, we say $V_i$ and $V_j$ in $U$ are sequentially-taken if they are neighbors in the taking sequence $U$ of rides. For example, when the UAV transfers from $V_i$ to $V_j$ in the sequence, we call $V_i$ and $V_j$ the prior and the following ride of the two sequentially-taken rides respectively.
\end{definition}
To verify whether two rides can be taken together by the UAV or not, we have Proposition \ref{compatibilitycondition}, which is summarized from (\ref{modelobjlimit001})-(\ref{model2const13}) and helps to determine whether two nodes/rides should be connected/compatible in the constructed graph.
\begin{proposition}[Compatible condition]\label{compatibilitycondition}
Given two rides $V_j$ and $V_i$ with $o_j<o_i$ and the power $P_j$ of the UAV at time $t_j$, we say they are compatible only when the UAV is able to take both  rides by transferring from $V_j$ to $V_i$. Specifically, they satisfy the following compatible condition (\ref{compatibleconditions}).
\begin{equation}\label{compatibleconditions}
    \begin{split}
&a) \;[{\rm time\;compatibility}] \;t_i\geq t_j+\frac{d_j-o_j}{v_j}+\frac{|o_i-d_j|}{v_0}  \\
&b)\; [{\rm power\; compatibility}]\;\textsc{PTF}(P_j,V_j,V_i)\geq 0
    \end{split}
\end{equation}
\end{proposition}
The moment while the UAV is either landing-on or flying-off a truck, note that the time and space dimensions keep in consistency between the UAV and the truck. Given two sequentially-taken rides $V_i$ (the prior one) and $V_j$, the UAV is supposed to land on $V_j$ right at time $t_j$ to catch the start of $V_j$ without reducing its remaining power at time $t_j$. This is because the \textit{power transfer function} in (\ref{powertransferfunctionnew}) is independent from when the UAV departs in transferring between $V_i$ and $V_j$. Or, the UAV can stop-and-recharge at the end location $d_i$ of $V_i$ until it can fly constantly to the start $o_j$ of $V_j$ (right at $t_j$). This helps us to better understand the location of the UAV while transferring between two rides. Theorem \ref{takensequence} shows the taking sequence of the UAV in a given set of accepted rides.

\begin{theorem}\label{takensequence}
Given the set $U=\{V_1,\cdots,V_{|U|}\}$ of rides that are accepted by the optimal solution (OPT), OPT takes all rides in $U$ following the increasing order of the rides' start locations, i.e. the smaller $o_i$ is, the earlier $V_i\in U$ is taken.
\end{theorem}

\textbf{Offline Algorithm}. At the high level, \textsc{OptimalHitching} first constructs a graph by screening unnecessary truck rides iteratively, which is on top of some characteristics of the problem. Afterwards, the optimal solution of the offline problem in this paper is converted to a minimum-weight shortest path of the graph. In the constructed graph, two virtual nodes $V_0$ and $V_{n+1}$ are introduced to represent the origin $O$ and the destination $A$ respectively, while the other nodes in $\{V_1,\cdots,V_n\}$ are constructed to represent the taking sequence of rides in $\mathbb{V}$ respectively due to Theorem \ref{takensequence}.
 Each node of $\{V_1,\cdots,V_n\}$ maintains a weight of the maximum power that the UAV could remain at the moment transferring to this node/ride, and connects to the previously constructed node from which the UAV transfers to the new node and remains that maximum power. In 
 other words, node weights in the graph are only used for checking the power compatibility of rides taken by the UAV. Due to
(\ref{arrivaltimelimittwo}), all edges of the graph are set as zero weight except for those connecting to $V_{n+1}$. In this way, the weight of a path connecting $V_0$/$O$ and $V_{n+1}$/$A$ indicates the arrival time of the UAV at $A$ taking those rides on the path. Below gives details of \textsc{OptimalHitching}.
 
 \begin{enumerate}
    \item \textit{First}, sort offline rides in $\mathbb{V}_n$, by increasing order of their start locations as $(V_1,V_2,\cdots,V_n)$; create virtual nodes $V_0=(0,0,0,0,0)$ and $V_{n+1}$ representing the origin $O$ and the destination $A$ respectively.
    \item  \textit{Then}, construct a graph $G=(N,E,w)$ with the weight function $w$ applying to both nodes in $N$ and edges in $E$, in an iterative way: 
    \begin{enumerate}
        \item  Include $V_0$ in $N$. Check in sequence $(V_1,V_2,\cdots,V_n)$ the rides one by one. When $V_i\in(V_1,V_2,\cdots,V_n)$ is compatible with at least one ride in $N$ by (\ref{compatibleconditions}), denote $\Phi(V_i)$ as the set of rides in $N$ that are compatible with $V_i$:
        \begin{enumerate}
            \item find in $\Phi(V_i)$ the node/ride, denoted by $V_*(i)$, to which the most power remains to the UAV on arrival at the start of $V_i$ by transferring from a ride in $\Phi(V_i)$;
            \item  include $V_{i}$ in $N$ and set weight $w(V_i)=\textsc{PTF}(w(V_*(i)),V_*(i),V_i)$; include ($V_*(i)$,$V_i$) in $E$ and set edge weight  $w(V_*(i),V_i)=0$; 
        \end{enumerate}
     \item Include $V_{n+1}$ in $N$. For each $V_j\in N$, add ($V_j,V_{n+1}$) to $E$ with weight $w(V_j,V_{n+1})=\xi(V_j)$ that representing the UAV's arrival time to $A$ with taking $V_j$ as the last ride. 
    \end{enumerate}
\item\textit{Finally}, find in $N$ the node $\overline{V}=\arg\min\limits_{V\in N-\{V_{n+1}\}}w(V,V_{n+1})$ that has the minimum-weight edge connecting with $V_{n+1}$, and further find backwards (from those nodes joining in $N$ earlier than $\overline{V}$) the node that connects with $\overline{V}$.\footnote{Step 2.b.\romannumeral1\; ensures only one node joining in $N$ earlier than $\overline{V}$ that connects with $\overline{V}$. } Repeat this step backwards until node $V_0$ is reached. Output the found nodes in the sequence of their joining time in $N$.
 \end{enumerate}
Note that the running time of \textsc{OptimalHitching} is dominated by graph construction steps, which is in $O(n^2)$. We have the following Theorem \ref{model2offlinethm1}. 

\begin{theorem}\label{model2offlinethm1}
\textsc{OptimalHitching} runs in $O(n^2)$-time and outputs the sequence of rides that are taken by an optimal offline solution.
\end{theorem}
\section{Lower Bounds on Competitive Ratios}\label{onlinesolutionswithcompetitiveanalysis}
\begin{table}
\caption{Derived notations for bound analysis}\label{derivednotations}
\begin{center}
\begin{tabular}{cp{180pt}}
\hline\hline
 Derived notations & Physical meanings\\
 \hline\hline
$T_{ru}:=\frac{\beta-\alpha}{\alpha\cdot v_0}$& the time duration for recharging the amount of power to fly a unit distance constantly.\\
\hline
$T_{mu}:=\frac{\beta}{\alpha\cdot v_0}$ &the time duration for moving a unit distance in the case that the UAV contains no power at the beginning.\\
\hline
$T_{f0}:=\frac{P_0}{\beta-\alpha}$ &the time duration that the UAV flies constantly from time $t_0$ on.\\
\hline
$l_{f}:=\frac{P_0v_0}{\beta-\alpha}$&the furthest location to which the UAV can reach by flying constantly from time $t_0$ on, note that $l_{f}=T_{f0}\cdot v_0$.\\
\hline
$\xi(\varnothing):=\frac{\beta}{\alpha}\frac{a}{v_0}-\frac{P_0}{\alpha}$ &the earliest arrival time of the UAV to $A$ when no ride is taken, by (\ref{arrivaltimelimittwo}) with $V_{|U|}=(0,0,0,0,v)$ and $P_{|U|}=P_0$.\\
\hline
$T_{ra}:=\frac{\beta-\alpha}{\alpha}\frac{a}{v_0}-\frac{P_0}{\alpha}$&the least time duration of the UAV to stop-and-recharge, to reach the target $A$ without taking rides, $T_{ra}=\xi(\varnothing)-\frac{a}{v_0}$.\\
\hline
 $L_{\rm min}:=\frac{(a\beta-a\alpha-P_0v_0)v}{v_0\alpha-v\alpha+v\beta}$& the minimum length to be saved by rides (with velocity $v$) to reach the target $A$ without stop-and-recharge, by  Lemma \ref{propforleastlength}.
\\
\hline
$Len(T):=\max\{\frac{(a\beta-a\alpha-P_0v_0-v_0T\alpha)v}{v_0\alpha-v\alpha+v\beta},0\}$& the least amount of length that the UAV needs to save for avoiding more time in stop-and-recharge afterwards, given that the UAV already stop-and-recharges for a total amount $T$ of time.\\
\hline\hline
\end{tabular}
\end{center}
\end{table}
We present the lower bounds of the UAV's travel time by first releasing a hook ride, and then releasing rides that are not compatible to ALG by power and time constraint if ALG rejects/accepts the hook ride, since taking a ride helps to save more energy in an early stage but moves more slowly. For bound analysis, we further derive some notations as summarized in Table \ref{derivednotations}.

Notice that the UAV does not need a ride $V_i$ with $\Delta t_i\geq T_{ra}+T_{f0}$. This is because the moment when $V_i$ starts, the UAV already contains at least $(\Delta t_i-T_{f0})\alpha$ of power by stop-and-recharging, which enables the UAV to fly constantly to $A$. Thus, we have the following proposition \ref{themaximumpossibletimegap}. Then, Lemma \ref{propforleastlength} is given to better figure out the minimum travel time that an OPT could achieve, based on which a lower bound is presented in Theorem \ref{genearllowerbound} as a benchmark for further online algorithm design.
\begin{proposition} \label{themaximumpossibletimegap}
$\Delta t_i=t_i-r_i$ of each ride $V_i\in \mathbb{V}$
ranges in $[0,T_{ra}+T_{f0}]$.
\end{proposition}

\begin{lemma}\label{propforleastlength}
The UAV needs to save an overall distance of at least $L_{\rm min}=\frac{(a\beta-a\alpha-P_0v_0)v}{v_0\alpha-v\alpha+v\beta}$ by taking rides, in order to avoid stop-and-recharge halfway.
\end{lemma}

\begin{theorem}\label{genearllowerbound}
For the problem with flexible $\Delta t_i\in [0,T_{ra}+T_{f0}]$, no online deterministic algorithm can achieve a competitive ratio better than 
\begin{equation}\label{generallbforlimit}
 \frac{\xi(\varnothing)-T_{mu}
 }{\frac{\left \lceil L_{\rm min}\right \rceil}{v}+\frac{a+1-\left \lceil L_{\rm min}\right \rceil}{v_0}}
\end{equation}
\end{theorem}
\section{Online Algorithms with Competitive Analysis}\label{onlinealgsection}
We propose a myopic algorithm \textsc{MyopicHitching} and a near-optimal algorithm $\Delta t$-\textsc{Adaptive} respectively, both of which inherit notations from Table \ref{derivednotations}. 
\subsection{\textsc{MyopicHitching} Algorithm} 
We first present the \textsc{MyopicHitching} algorithm under fixed $\Delta t$. By some small changes in the following accepting conditions (\romannumeral1-\romannumeral2), one can easily extend it to the flexible $\Delta t$. \textsc{MyopicHitching} follows the route by default to fly forward constantly until the first time it runs out of  power. Afterwards, the UAV stops-and-recharge until containing enough power to fly constantly to the target $A$. The by-default action possibly changes only when a ride is accepted. Denote $P(V)$ as the power that the UAV remains at the start time of an accepted ride $V$ in the current solution, and $U$ as the set of rides accepted by \textsc{MyopicHitching}. A new ride $V_i\in\mathbb{V}$ is accepted only when $V_i$ meets the following two conditions together, i.e., $l_{\rm rc}\cdot l_{\rm aa}=1$. Accordingly, update both $P(V_i)$ and power attributes of those rides in $U$ that depart after $V_i$ by the power-transfer-function in (\ref{powertransferfunctionnew}): 

\textbf{\romannumeral1) }\textit{ride-compatible}\footnote{Under fixed $\Delta t$, online algorithm must accept ride that depart after previously accepted rides due to Observation \ref{earlyreleaseearlydepart}. But this is not the case under flexible $\Delta t$, a newly accepted ride can depart between two rides in $U$, and the potential $V_{\rm right}=\arg\min\limits_{\mathclap{\{V_j\in U| o_j>o_i\}}}o_j-o_i$ should also be compatible with $V_i$.}: the new ride should meet (\ref{compatibleconditions}) with $V_{\rm left}= \arg\min\limits_{\mathclap{\{V_j\in U| o_j<o_i\}}}\{o_i-o_j\}$ which is the only  ride in $U$ that is sequentially-taken with $V_i$, i.e., the following indicator should be equal to 1.
$$\mathbf{\textup{1}}_{\rm rc}=\left\{\begin{matrix}
1, &\begin{matrix}
t_i\geq t_{\rm left}+\frac{d_{\rm left}-o_{\rm left}}{v_{\rm left}}+\frac{|o_i-d_{\rm left}|}{v_0} \\ {\rm \;\textbf{and}\;}\textsc{PTF}(P_{\rm left},V_{\rm left},V_i)\geq 0\\ 
\end{matrix}\\ 
 0,&{\rm  otherwise}
\end{matrix}\right.$$

\textbf{\romannumeral2)} \textit{arrival-ahead}: the UAV will reduce its overall travel time when taking the new ride, i.e., the following indicator which is due to (\ref{arrivaltimelimit}) should be equal to 1.
\begin{center}
$\mathbf{\textup{1}}_{\rm aa}=\left\{\begin{matrix}
1, &\xi(U\cup \{V_i\})\leq\xi(U)\\ 
 0,&{\rm  otherwise}
\end{matrix}\right.$ \end{center}

To take each accepted ride in $U$, \textsc{MyopicHitching} guides the UAV to reach the origin of the ride right at its start time. Whenever the UAV contains enough power to fly constantly towards the target $A$, i.e., (\ref{powerenoughconditionalg}) is satisfied, it stops accepting new ride and flies directly to $A$ after taking the last-accepted ride. 
\begin{equation}\label{powerenoughconditionalg}
  \left\{\begin{matrix}
u_{\textup{time}}\geq\max\{(\frac{\beta}{\alpha}-1)\frac{a-d_{|U|}}{v_0}-\frac{\widetilde{P}(V_{|U|})}{\alpha}, t_{|U|}+\frac{d_{|U|}-o_{|U|}}{v_{|U|}}\}{\rm\;or}\\ 
\widetilde{P}(V_{|U|})\geq \frac{a-d_{|U|}}{v_0}(\beta-\alpha) {\rm \;\&\;} u_{\textup{time}}<t_{|U|}+\frac{d_{|U|}-o_{|U|}}{v_{|U|}}{\rm\;or}\\
U=\varnothing {\rm \;\&\;} u_{\textup{time}}=T_{ra}
\end{matrix}\right.  
\end{equation}
in which $\widetilde{P}(V_{|U|})=P(V_{|U|})+\frac{\alpha(d_{|U|}-o_{|U|})}{v_{|U|}}$ indicates the power remaining to the UAV on completing $V_{|U|}$, while $u_{\rm time}$ indicates the real time in the execution.

\begin{observation}\label{earlyreleaseearlydepart}
For the problem with fixed $\Delta t$, the earlier a ride is released, the earlier the ride departs.
\end{observation}
\begin{lemma}
For the problem with fixed $\Delta t$, \textsc{MyopicHitching} always takes rides in increasing order of their start locations.
\end{lemma}
Notice in the following example that \textsc{MyopicHitching} has a \textit{defect} which can be exploited by an adversary leading to a very bad competitive ratio: suppose $V_1=(\frac{1}{2v_0},\frac{1}{2v_0}+\Delta t, \varepsilon,1+\varepsilon,v)$ is the first released ride with a small $\Delta t$. At the release time $\frac{1}{2v_0}$ of $V_1$, the UAV is at location $\frac{1}{2}$ and contains power $\overline{P}=P_0+(\alpha-\beta)\frac{1}{2v_0}$.
We note that the \textit{arrival-ahead} condition (\romannumeral2) implies a ride released at an early stage could be accepted when the ride could help the UAV to save more energy. Since the UAV can save a small amount $\frac{2\varepsilon}{v_0}\beta$ of power by taking $V_1$, $V_1$ is accepted by accepting conditions (\romannumeral1)-(\romannumeral2). Notice that \textsc{MyopicHitching} actually costs
$t_{\rm wait}=\Delta t +\frac{1}{v_1}+\frac{\frac{1}{v_0}-1-2\varepsilon}{v_0}$ of waiting time. The \textit{defect} appears when $\Delta t<t_{\rm wait}$ since the adversary could further releases rides making \textsc{MyopicHitching} violate constraints in (\ref{compatibleconditions}).  When $\varepsilon\rightarrow 0$, we get the following Theorem \ref{competitveratio1}.

\begin{theorem}\label{competitveratio1}
For the problem with fixed $\Delta t$,  \textsc{MyopicHitching} achieves a competitive ratio no worse than
\begin{equation}\label{onlinealgorithm1ratio}
\frac{\xi(\varnothing)}{(a-\left \lceil  {\rm L}_{\rm min}\right \rceil)T_{mu}-\frac{P_0}{\alpha}}
\end{equation}
\end{theorem}
Particularly, in the scenario where the UAV already exhausts its power (i.e., $P_0=0$) at the very beginning, \textsc{MyopicHitching} achieves a competitive ratio no worse than $\frac{a}{a-\left \lceil  {\rm L}_{\rm min}\right \rceil}$.
\subsection{$\Delta t$-\textsc{Adaptive} Algorithm}
We fix  the \textit{defect} of \textsc{MyopicHitching} by leveraging adaptability of $\Delta t$ in \textsc{MyopicHitching} and present $\Delta t$-\textsc{Adaptive} algorithm  by including the following \textit{conditional-start} condition (i.e., $\textup{1}_{\rm cs}=1$) in the ride-accepting conditions. 
Specifically, $\Delta t$-\textsc{Adaptive} keeps the \textbf{Input, Lines 1-4, 6-15} of the pseudocode in Appendix \ref{APPENDIXMYOPIC} of \textsc{MyopicHitching} the same, but replaces the \textbf{if condition} $\textup{1}_{\rm rc}\cdot \textup{1}_{\rm aa}$==1 in \textbf{Line 5} by $\textup{1}_{\rm rc}\cdot \textup{1}_{\rm aa}\cdot\textup{1}_{\rm cs}$==1.

\textbf{\romannumeral3}): \textit{conditional-start} condition: the following indicator $l_{\rm cs}=1$.
$$\textup{1}_{\rm cs}=\left\{\begin{matrix}
1, & \begin{matrix}
o_i\geq \frac{\Delta t v_0}{2}+l(u_{\rm time}){\rm \;for\;} u_{\rm time}\in[0,\frac{l_f}{v_0}],\\ 
 {\rm or,\;} o_i\geq l_f+\frac{\Delta t v_0}{2}{\rm \;for\;}u_{\rm time}\in(\frac{l_f}{v_0},\xi(\varnothing)]
\end{matrix}\\ 
0, & \textup{otherwise}
\end{matrix}\right.$$

\begin{theorem}\label{ratioforalgorithm3}
For the problem with fixed $\Delta t$, $\Delta t$-\textsc{Adaptive} algorithm achieves a competitive ratio no worse than
\begin{equation}\label{upperboundforalg3}
    \frac{\xi(\varnothing)}{\frac{\Delta t}{2}+\frac{a-Len(\frac{\Delta t}{2})}{v_0}+\frac{Len(\frac{\Delta t}{2})}{v}}
\end{equation}
\end{theorem}

Recall that rides with $\Delta t\geq T_{ra}+T_{f0}$ are not worth taking (see Lemma \ref{themaximumpossibletimegap}). By Theorem \ref{ratioforalgorithm3}, we know $\Delta t$-\textsc{Adaptive} prefers larger $\Delta t$ since its competitive ratio decreases as $\Delta_t$ increases. Comparing (\ref{upperboundforalg3}) and (\ref{onlinealgorithm1ratio}), we find that $\Delta t$-\textsc{Adaptive} algorithm outperforms \textit{MyopicHitching} especially when $\Delta t$ is large, this is because $\Delta t$-\textsc{Adaptive} guarantees that OPT has to cost some waiting time of at least $\frac{\Delta t}{2}$ in the case when all released truck rides are not compatible to $\Delta t$-\textsc{Adaptive}. What's more, $\Delta t$-\textsc{Adaptive} algorithm actually achieves near-optimal performance compared to the best possible online algorithm since the latter can save at most one ride while the OPT has to pay a waiting time of $\Delta t$. Please refer to the figure in the appendix \ref{comparisonfigure} for some comparison between \textit{MyopicHitching} and $\Delta t$-\textsc{Adaptive}.
\section{Concluding Remarks}
In this paper, we give the first theoretical work on the problem of online ride-hitching in UAV travelling.  By mapping truck trips to nodes in a graph in an iterative way, we give a shortest-path-like solution for the offline version of this problem where truck trips are all known in advance. As a benchmark, we present lower bounds on the competitive ratio of the problem, respectively, for different settings. Then, we show that a greedy algorithm which accepts as many rides as possible has a defect. To fix the defect, we propose the $\Delta t$-\textsc{Adaptive} algorithm, achieving near-optimal performance in terms of the competitive ratio. 

\textbf{Acknowledgement.} We thank the anonymous referees for their helpful feedback. Part of this work was done while Songhua LI was visiting the Singapore University of Technology and Design. Minming Li is also from City University of Hong Kong Shenzhen Research Institute, Shenzhen, P.R. China. The work described in this paper was sponsored by Project 11771365 supported by NSFC.

\newpage
\appendix
\section{Proof of Proposition \ref{objinlimittransfer}}
\begin{proof}
Given the ride  $V_{|U|}=(r_{|U|},t_{|U|},o_{|U|},d_{|U|})$ that is finally taken by the UAV among those taken rides in $U$, the overall flying distance of the UAV, denoted by $dist_f$, is 
\begin{equation}
dist_f=o_1+\sum_{i=1}^{|U|-1} |o_{i+1}-d_i|+a-d_{|U|}
\end{equation}
We also note that the UAV flies off $V_{|U|}$ at time $t_{|U|}+\frac{d_{|U|}-o_{|U|}}{v_{|U|}}$. Denote $P_{\rm r}$ as the power remains to the UAV at time $t_{|U|}+\frac{d_{|U|}-o_{|U|}}{v_{|U|}}$, we discuss two cases. \\
\textbf{Case 1.} $P_{\rm r}\geq \frac{a-d_{|U|}}{v_0}(\beta-\alpha)$. The UAV contains enough power to fly constantly from $d_{|U|}$ (the fly-off point of $V_{|U|}$) to the target $A$. Then, the arrival time $\xi(U)$ to $A$, which corresponds to $V_{|U|}$ only, is $t_{|U|}+\frac{d_{|U|}-o_{|U|}}{v_{|U|}}+\frac{a-d_{|U|}}{v_0}$. 
Meanwhile, we have  $\xi(U)\geq \frac{dist_f}{v_0}\frac{\beta}{\alpha}-\frac{P_0}{\alpha}$ in this case, 
since the UAV still contains non-negative power on arrival at $A$ at time $\xi(U)$, i.e.,  $P_0+\xi(U)\alpha-\frac{dist_f}{v_0}\beta\geq 0$.
\\
\textbf{Case 2.} $P_{\rm r}< \frac{a-d_{|U|}}{v_0}(\beta-\alpha)$.  The UAV does not have enough power to fly constantly from $d_{|U|}$ to the target $A$. To achieve the earliest possible arrival time to $A$, the UAV needs to stop-and-recharge for an amount of power at least $\frac{a-d_{|U|}}{v_0}(\beta-\alpha)-P_{\rm r}$ (and runs out of its power on arrival at $A$). This implies $\xi(U)=\frac{dist_f}{v_0}\frac{\beta}{\alpha }-\frac{P_0}{\alpha}$ in this case. Clearly, $\xi(U)>t_{|U|}+\frac{d_{|U|}-o_{|U|}}{v_{|U|}}+\frac{a-d_{|U|}}{v_0}$ as otherwise it contradicts with the base of this case.
\end{proof}
\section{Proof of Theorem \ref{takensequence}}
\begin{proof}
Clearly, all rides accepted by OPT are finally taken. We just need to show that any two adjacent rides in $U$ are actually sequentially-taken with the same order by OPT. For the sake of contradiction, suppose there are two adjacent rides $V_i$ (the prior one) and $V_j$ that are not sequentially-taken, i.e., the UAV takes $V_j$ (with landing-on location $o_j$) prior to $V_i$ (with the landing-on location $o_i$). Notice that, the UAV can save $(\beta-\alpha)\frac{d_j-o_i}{v_0}$ more amount of energy by taking $V_i$ first since $o_i<o_j$, which either contradicts with OPT (in the case that the saved amount of energy leads to earlier arrival time to $A$) or generates another optimal solution (in the case that this saved amount of energy does not induce earlier arrival time to $A$), which is because the land-on point (which is the start location of the ride) of each taken ride is unique. Therefore, any two adjacent rides in OPT is sequentially-taken.
\end{proof}
\section{Proof of Lemma \ref{propforleastlength}}
\begin{proof}
Suppose, w.l.o.g., that the UAV takes rides in $\{V_1,\cdots, V_m\}$  to assist itself to reach the target $A$ without stopping halfway. Denote $L$ as the overall length that the UAV saves by taking rides. As the UAV does not stop halfway, the final arrival time of the UAV at the destination $A$, denoted by $\textup{t}_{\rm final}$, consists of two parts, the time spent on rides (denoted by $\textup{t}_{\rm r}$) and the flying time $\frac{a-\textup{L}}{v_0}$. Denote $\textup{L}_1,\cdots,\textup{L}_m$ as the accumulative length in $L$ by rides $V_1,\cdots, V_m$, respectively. Then, $L=\sum_{i=1}^{i=m}\textup{L}_i$ and $\textup{t}_{\rm r}=\frac{\textup{L}_1}{v_1}+\cdots+\frac{\textup{L}_m}{v_m}$. Further, 
\begin{equation}\label{leastlength01}
    \textup{t}_{\rm final}=\frac{\textup{L}_1}{v_1}+\cdots+\frac{\textup{L}_m}{v_m}+\frac{a-L}{v_0}\leq \frac{L}{v}+\frac{a-L}{v_0}
\end{equation}
in which the inequality holds by $v=\min\limits_{\mathclap{i\in\{1,2,\cdots,m\}}}v_i$. The moment when UAV just reaches the target $A$, note that the power of the UAV remains non-negative. Hence,
\begin{equation}\label{leastlength02}
    P_0+\alpha \textup{t}_{\rm final}-\beta(\frac{a-L}{v_0})\geq 0
\end{equation}
Substituting (\ref{leastlength01}) in (\ref{leastlength02}), we have
\begin{equation*}
L\geq \frac{a(\beta-\alpha)-P_0v_0}{v_0}\cdot \frac{vv_0}{v_0\alpha+(\beta-\alpha)v}=\frac{(a\beta-a\alpha-P_0v_0)v}{v_0\alpha-v\alpha+v\beta}
\end{equation*}
\end{proof}
\section{Omitted Proof of Theorem \ref{genearllowerbound}}
\begin{proof}
Suppose all rides move with constant velocity $v$ and are of unit-length. By the basic settings $P_0<(\beta-\alpha)\frac{a}{v_0}$ and $v_0>v$ of this paper, we have $L_{\rm min}>0$ and further  $\left \lceil  L_{\rm min} \right \rceil\geq 1$. Define $\tau=\left \lceil  L_{\rm min} \right \rceil$ and $\varsigma =\left \lfloor l_f \right \rfloor$. Suppose the length $a$ of the trip $OA$ is larger than $l_f+\tau$. Now, we consider the first released ride $V_1=(0,T_{f0},l_f,l_f+1,v)$ which can only be taken when the UAV departs from $t_0$ and consumes its entire power on arrival at the start location $l_f$, i.e., the orange rectangle/ride in Fig. \ref{lb_case2}. 
\\
\textbf{Case 1.} ALG accepts $V_1$. We discuss two sub-cases.
\\
\textbf{Case 1.1.} $\tau\geq \varsigma+2$. After a short time duration $\varepsilon$ ($\geq\frac{1}{v_0}$), the adversary further releases another $\varsigma$ head-to-tail rides within $[0,L_{\rm min}]$: for $i\in\{2,\cdots,\varsigma+1\}$
\begin{equation}
V_i=(\varepsilon+\frac{i-2}{v},\varepsilon+\frac{i-2}{v},i-2,i-1,v)
\end{equation}
Notice that, at time $\varepsilon$ when $V_2$ departs, ALG already flies over the location $\frac{1}{v_0}v_0=1$, implying that ALG does not need to accept $V_2$. In other words, ALG does not fly backward until reaching the origin of $V_1$. 
Note that the distance between the end location $\varsigma$ of ride $V_{\varsigma+1}$ with the location $L_{\rm min}$ is smaller than 1, which takes $T_{f0}-\frac{\varsigma}{v_0}$ of time for flying. Further at time $\varepsilon +\frac{\varsigma}{v}+T_{f0}-\frac{\varsigma }{v_0}$ (denoted by $\textup{t}_{1}$), when OPT just flies-off $V_{\varsigma+2}$, the adversary releases $V_{\varsigma+2}=(\textup{t}_{1},\textup{t}_{1},l_f,l_f+1,v)$. Clearly, the above $\varsigma+1$ rides $\{V_2,\cdots,V_{\varsigma+2}\}$ cannot be taken by ALG since they are not compatible with $V_1$.  However, OPT can successfully take all rides in $\{V_2,\cdots,V_{\varsigma+2}\}$ without violating the compatible constraints in (\ref{compatibleconditions}), see the blue rectangles/rides in Fig. \ref{lb_case2}. At time $\textup{t}_{1}+\frac{1}{v}$, when completing taking the rides and flying-off $V_{\varsigma+2}$, OPT is located at ($l_f+1$) and contains power of

$\begin{aligned}
P_{\rm OPT}&=P_0+\underbrace{(\varepsilon +\frac{\varsigma +1}{v}-\frac{\varsigma }{v_0}+T_{f0})\alpha}_{\rm recharging\;power}-\frac{l_f-\varsigma }{v_0}\beta\\
&=P_0+(\varepsilon +\frac{\varsigma +1}{v})\alpha+(l_f-\frac{\varsigma }{v_0})(\alpha-\beta)
\end{aligned}$
\\
Meanwhile at time $\textup{t}_{1}+\frac{1}{v}$, ALG is also located at ($l_f+1$) but contains power of just $P_{\rm ALG}=(\varepsilon +\frac{\varsigma +1}{v}-\frac{\varsigma }{v_0})\alpha$, since ALG just runs out of its power at time $T_0^f$ on arrival at $l_f$. Right at time $\textup{t}_{1}+\frac{1}{v}$, the adversary further releases another $\tau-\varsigma-2$ head-to-tail rides, i.e., 
\begin{center}
$V_{\varsigma+3}=(\textup{t}_{1}+\frac{1}{v},\textup{t}_{1}+\frac{1}{v}+\frac{P_{\rm OPT}}{\beta-\alpha},l_f+1+\frac{P_{\rm OPT}v_0}{\beta-\alpha},l_f+2+\frac{P_{\rm OPT}v_0}{\beta-\alpha},v)$
    
, $\cdots$,
    
$ V_{\tau+1}=(\textup{t}_{1}+\frac{1}{v}+\frac{\tau-\varsigma-3}{v},\textup{t}_{1}+\frac{1}{v}+\frac{P_{\rm OPT}}{\beta-\alpha}+\frac{\tau-\varsigma-3}{v},l_f+\tau-\varsigma-2+\frac{P_{\rm OPT}v_0}{\beta-\alpha},l_f+\tau-\varsigma-1+\frac{P_{\rm OPT}v_0}{\beta-\alpha},v)$
\end{center}
Note that OPT can just catch $V_{\varsigma+3}$ by flying from $l_f+1$ at the time $\textup{t}_{1}+\frac{1}{v}$ when UAV just takes off from $V_{\varsigma+2}$ with its entire power. This implies that the OPT can take all rides in $\{V_{\varsigma+4},\cdots,V_{\tau+1}\}$ successively (see those blue rectangles/rides in Fig. \ref{lb_case2}). Hence, OPT can reach the target $A$ without stop-and-recharge halfway by riding $\{V_2,\cdots,V_{\tau+1}\}$ (see Lemma \ref{propforleastlength}).  As OPT only needs to stop-to-recharge some time of $\varepsilon$ from the very beginning to wait for the departure of $V_2$, the arrival time of OPT to the target $A$ is $\xi_{\rm OPT}=\varepsilon+\frac{\tau}{v}+\frac{a-\tau}{v_0}$, which includes the time $\frac{\tau}{v}$ spent on rides and the time 
$\frac{a-\tau}{v_0}$ while flying. In contrast, the ALG cannot even catch up $V_{\varsigma+3}$ as the power constraint is violated by $P_{\rm ALG}<P_{\rm OPT}$, and further ALG cannot take rides in $\{V_{\varsigma+3},\cdots,V_{\tau}\}$. In other words, ALG can only fly by itself towards $A$ after taking $V_1$. 
Hence, the arrival time of ALG is $\xi_{\rm ALG}\geq T_{f0}+(a-1-l_f)\cdot\frac{\beta}{\alpha v_0}$.  When $\varepsilon \rightarrow \frac{1}{v_0}$, we get
\begin{equation}
\rho=\frac{\xi_{\rm ALG}}{\xi_{\rm OPT}}\geq \frac{ T_{f0}+(a-1-l_f)\cdot\frac{\beta}{\alpha v_0}}{\varepsilon +\frac{\tau}{v}+\frac{a-\tau}{v_0}}\rightarrow\frac{(a-1)\frac{\beta}{\alpha v_0}-\frac{P_0}{\alpha}}{\frac{\tau}{v}+\frac{a+1-\tau}{v_0}}\end{equation}
\\
\textbf{Case 1.2.} $\tau\leq  \varsigma+1$. After time $\varepsilon$, the adversary releases $\tau$ rides sequentially, including  head-to-tail rides
$V_i=(\varepsilon+\frac{i-2}{v},\varepsilon+\frac{i-2}{v},i-2,i-1,v)$ for $i\in\{2,\cdots,\tau\}$  and $V_{\tau+1}=(\varepsilon+\frac{\tau-1}{v},\varepsilon+\frac{\tau-1}{v},l_f,l_f+1,v)$.  OPT can take rides in $\{V_2,\cdots,V_{\tau+1}\}$ without violating compatible constraint. By a similar analysis as in subcase 1.1,  we have a lower bound of this subcase as  $\rho=\frac{\xi_{\rm ALG}}{\xi_{\rm OPT}}\geq \frac{(a-1)\frac{\beta}{\alpha v_0}-\frac{P_0}{\alpha}}{\frac{\tau}{v}+\frac{a+1-\tau}{v_0}}$.
\\
\textbf{Case 2.} ALG rejects $V_1$ at $t_0$.  Note that the earliest arrival time (say $\textup{t}_a$) of ALG to $l_f+1$ is  $T_{f0}+\frac{\beta}{\alpha v_0}$, which is derived by $P_0+\textup{t}_a\alpha-\beta(T_{f0}+\frac{1}{v_0})\geq 0$. This implies ALG should contain non-negative power at $\textup{t}_a$. Later, the adversary only releases rides
with start locations larger than $l_f+1$ and start time earlier than time $T_{f0}+\frac{\beta}{\alpha v_0}$, which are compatible in OPT but not in ALG. For example, rides in Fig. \ref{lb_case1}. 
By a similar idea in Case 1, one can derive a lower bound of this case as $\rho=\frac{\xi_{\rm ALG}}{\xi_{\rm OPT}}\geq \frac{T_{f0}+\frac{\beta}{\alpha\cdot v_0}(a-l_f)}{\frac{\tau}{v}+\frac{a-\tau}{v_0}}=\frac{a\frac{\beta}{\alpha v_0}-\frac{P_0}{\alpha}}{\frac{\tau}{v}+\frac{a-\tau}{v_0}}$.

Therefore, the lower bound could be got as  summarized in (\ref{generallbforlimit}).
\begin{landscape}
\begin{figure}
    \centering
    \includegraphics[width=15cm,height=10cm]{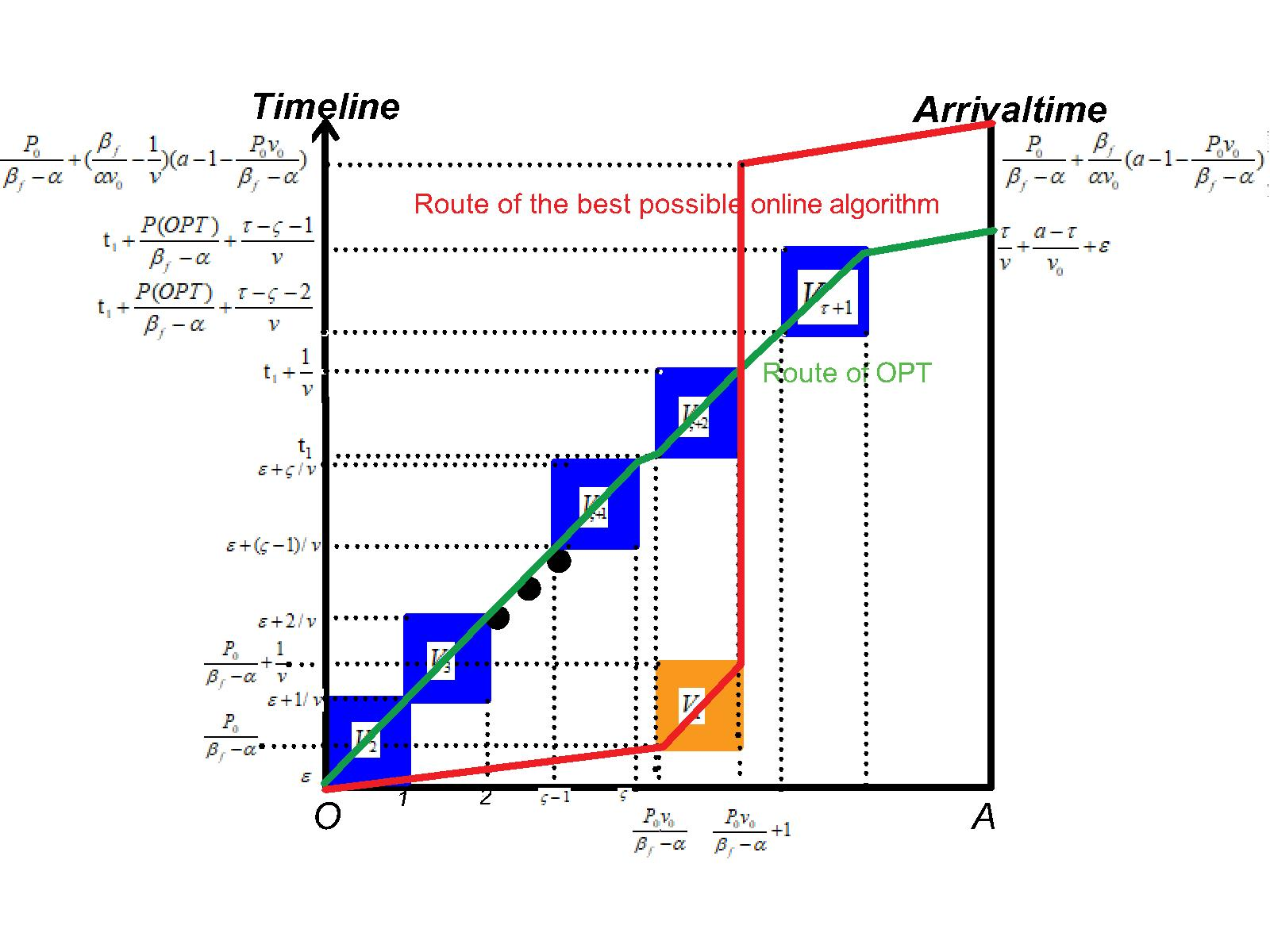}
    \caption{Illustration example, in which the left-side, right-side, bottom-side and top-side of a rectangle represent the start location, end location, start time, and the end time of the ride, respectively. In this case, ALG accepts the hook ride (which is the isolated orange rectangle), then, the adversary releases rides before the hook ride (not compatible with the hook ride) making ALG stores less energy than OPT does after the hook ride. Further, more rides can be released that are not compatible in ALG due to constraint (\ref{compatibleconditions}).}
    \label{lb_case2}
\end{figure}
\end{landscape}
\begin{landscape}
\begin{figure}
    \centering
    \setlength{\abovecaptionskip}{0.cm}
    \includegraphics[width=15cm,height=10cm]{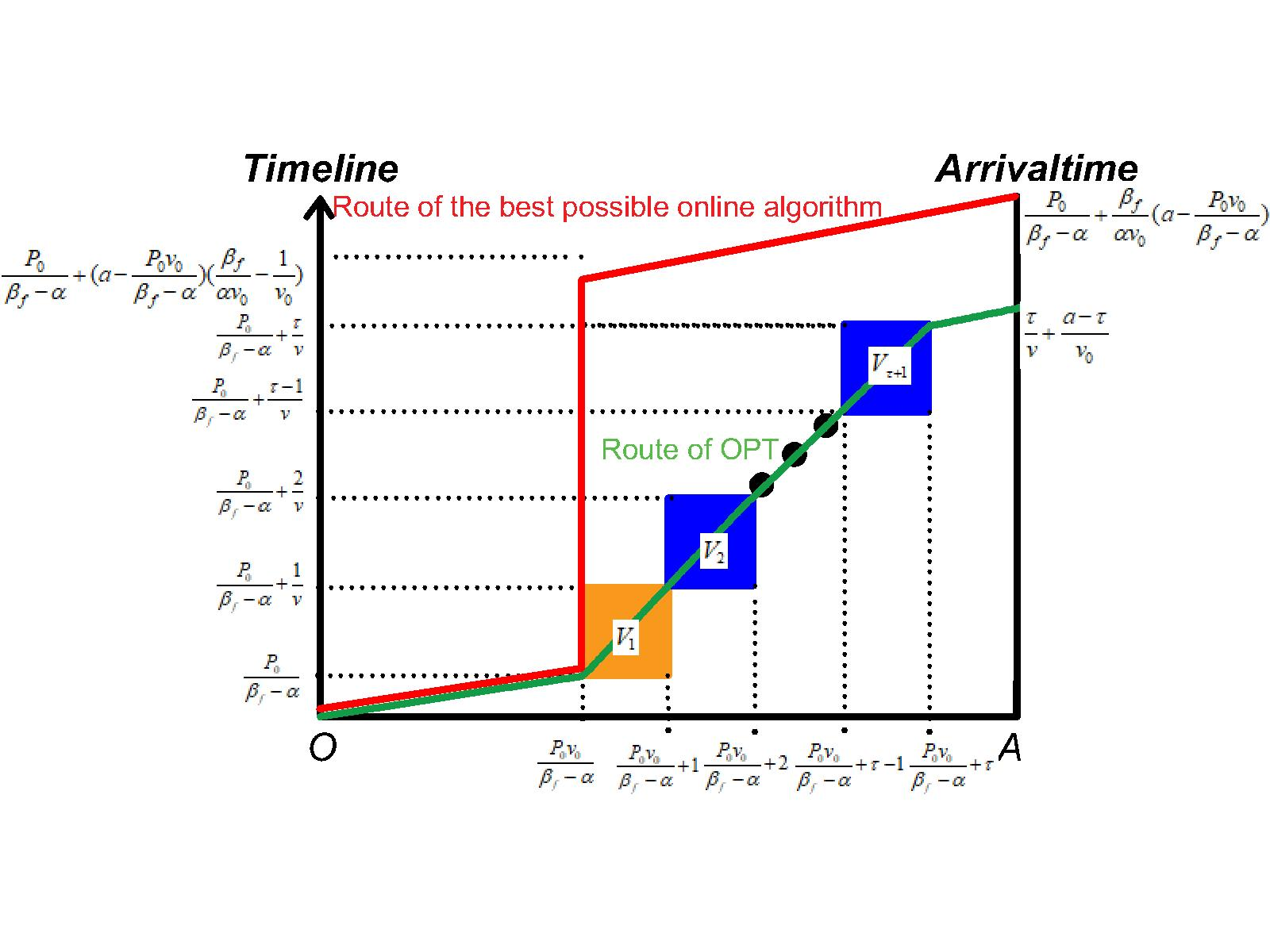}
    \caption{Illustration of rides configurations in Case 2. The ALG rejects the hook ride (the left-most one in orange of  Fig. \ref{lb_case1}), then, the adversary only releases rides after the hook ride making ALG not able to take as violate constraint (\ref{compatibleconditions}) in power.}
    \label{lb_case1}
\end{figure}
\end{landscape}
\end{proof}
\section{Omitted Proof of Theorem 5}
\begin{proof}
Our discussion centers around the ride, denoted by $\overline{V}=(\overline{r},\overline{r}+\Delta t,\overline{o},\overline{d},\overline{v})$, which is the first ride accepted by either OPT or ALG but not both. By Observation \ref{earlyreleaseearlydepart}, $\overline{V}$ is also the ride that is firstly taken by either OPT or ALG. Denote $l(\overline{r})$ as the real-time location of ALG. 
\\
\textbf{Case 1.} $\frac{\Delta t}{2}>\frac{l_f}{v_0}$ and $\overline{r}\in [0,\frac{l_f}{v_0}]$. 
\\
\textbf{Case 1.1.} $V$ is accepted by ALG. This implies $\xi_{\rm ALG}<\xi(\varnothing)$ as otherwise $V$ is rejected by the \textit{arrival-ahead condition} $\textup{1}_{\rm aa}$, and $\overline{o}\geq
\frac{\Delta t v_0}{2}+l(\overline{r})\geq \frac{\Delta t v_0}{2}>l_f$. Hence, ALG stops-and-recharge some time of 
$\Delta t-\frac{l_f-l(\overline{r})}{v_0}-\frac{\overline{o}-l(\overline{r})}{v_0}=\Delta t+\frac{2l(\overline{r})-\overline{o}-l_f}{v_0}$ (denoted by ${\rm T}_{\rm ALG}$).
Later, one has to stop-and-recharge some time more than ${\rm T}_{\rm ALG}$ since $\overline{r}$ increases. Suppose OPT totally accepts $x$ ($\geq 2$) rides, note that OPT stops-and-recharge at least $\frac{\Delta t}{2}$ for each ride after the first ride  since $\beta\geq 2\alpha$. Hence, we get
\begin{equation}
\begin{aligned}
\rho&\leq \frac{\xi(\varnothing)}{\frac{(x-1)\Delta t}{2}+\frac{a-Len(\frac{(x-1)\Delta t}{2})}{v_0}+\frac{Len(\frac{(x-1)\Delta t}{2})}{\overline{v}}}\\
&\leq \frac{\frac{\beta}{\alpha}\frac{a}{v_0}-\frac{P_0}{\alpha}}{\frac{\Delta t}{2}+\frac{a-Len(\frac{\Delta t}{2})}{v_0}+\frac{Len(\frac{\Delta t}{2})}{\overline{v}}}\\
&\leq   \frac{\xi(\varnothing)}{\frac{\Delta t}{2}+\frac{a-Len(\frac{\Delta t}{2})}{v_0}+\frac{Len(\frac{\Delta t}{2})}{v}}
\end{aligned}
\end{equation}
\\
\textbf{Case 1.2.} $V$ is accepted by OPT. This infers that the OPT has to stop-and-recharge for some time at least $\frac{\Delta t}{2}$, implying

$$\rho=\frac{\xi_{\rm ALG}}{\xi_{\rm OPT}}\leq    \frac{\xi(\varnothing)}{\frac{\Delta t}{2}+\frac{a-Len(\frac{\Delta t}{2})}{v_0}+\frac{Len(\frac{\Delta t}{2})}{v}}$$
\\
\textbf{Case 2.} $\frac{\Delta t}{2}>\frac{l_f}{v_0}$ and $\overline{r}\in (\frac{l_f}{v_0},\xi(\varnothing)]$, or  $\frac{\Delta t}{2}\leq \frac{l_f}{v_0}$. By applying a similar idea as Case 1, one can derive an upper bound of $   \frac{\xi(\varnothing)}{\frac{\Delta t}{2}+\frac{a-Len(\frac{\Delta t}{2})}{v_0}+\frac{Len(\frac{\Delta t}{2})}{v}}$ as well. 

\end{proof}

\section{Pseudocode of \textsc{OptimalHitching} Refers to Algorithm \ref{offlineALG_model2variant}}

\begin{algorithm}[tb] 
\caption{\textsc{OptimalHitching}}
\label{offlineALG_model2variant}
\textbf{Input}: A set $\mathbb{V}_n$ of rides, $\beta$, $\alpha$, $v_o$, $P_0$, $a$;\\
\textbf{Output}: A set $U$ of accepted rides;\\
\begin{algorithmic}[1] 
\STATE Initialization. $V_0\leftarrow (0,0,0,0,0)$, $i\leftarrow 1$, $N\leftarrow \{V_0\}$, $E\leftarrow \varnothing$, $w(V_0)\leftarrow P_0$;
\STATE Sort rides in $\mathbb{V}_n\cap \{V_0\}$ in increasing order of their origins, as $(V_0,V_1,V_2,\cdots,V_n)$;
\FOR {$i\leq n$}
\STATE $\Phi(i)\leftarrow \{V_j\in N|V_j {\rm \;and\;} V_i {\rm\;are\; compatible\;in\;(\ref{compatibleconditions})}\}$;
\IF {$\Phi(i)==\varnothing$}
\STATE $i++$;
\ELSE
\STATE $V_*(i)\leftarrow \arg\max\limits_{V_j\in \Phi(i)}\{\textsc{PTF}(w(V_j),V_j,V_i)\}$;
\STATE  $N\leftarrow N\cup \{V_*(i)\}$, $E\leftarrow E\cup\{(V_*(i),V_i)\}$, $w(V_i)\leftarrow \textsc{PTF}(w(V_*(i)),V_*(i),V_i)$;\\
$w(V_*(i),V_i)\leftarrow 0$, $i++$;
\ENDIF
\ENDFOR
\STATE $N\leftarrow N\cup \{V_{n+1}\}$;
\FOR{$V_j\in N-\{V_{n+1}\}$}
\STATE  $E\leftarrow E\cup \{(V_j,V_{n+1})\}$, $w(V_j,V_{n+1})\leftarrow  \xi(V_{j})$ by (\ref{arrivaltimelimittwo});
\ENDFOR
\STATE Output the nodes on the minimum-weight path connecting $V_0$ and $V_{n+1}$ in $G=(N,E,w)$.
\end{algorithmic}
\end{algorithm}
\newpage
\section{Comparison Figure}\label{comparisonfigure}
\begin{figure}
    \centering
    \includegraphics[width=8cm,height=5cm]{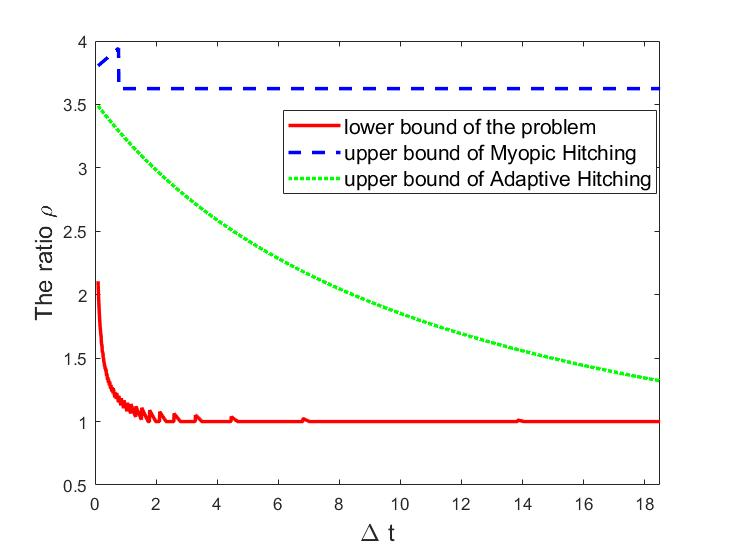}
    \caption{Performance comparison in algorithms with input parameters normalized as $a$=100 (km), $V_0$=100 (km/h),  $V=60$ (km/h), $\frac{\beta}{\alpha}$= 6, $P_0$ = 20.}
    \label{performancecomparison}
\end{figure}
\section{Pseudocode of \textsc{MyopicHitching} Refers to Algorithm \ref{simplemyopicalg}}\label{APPENDIXMYOPIC}
\begin{algorithm}
\caption{\textsc{MyopicHitching}}
\label{simplemyopicalg}
\textbf{Input}:  $\mathbb{V}_n=\{V_1,\cdots,V_n\}$, $\beta$, $\alpha$, $v_o$, $P_0$, $OA=a$;
\\\textbf{Output}: $U\subseteq \mathbb{V}_n$;\\
\begin{algorithmic}[1] 
\STATE \textit{Initialization.} $U\leftarrow \varnothing$, $u_{\textup{time}}\leftarrow t_0$;\quad\quad\COMMENT{\%$u_{\textup{time}}$: real time}\\
\STATE \textit{Route by default when no ride to be taken}. \\Before the first time using up power, UAV flies over $u_{\textup{time}}$ with velocity $v_0$. Afterwards, the UAV does not move until it contains enough power to fly constantly to the target $A$.\\
\COMMENT{\% the following loop shows \textit{ride-accepting procedure}}\\
\WHILE{$u_{\textup{time}}==r_i$}
\STATE  $V_{\rm left}\leftarrow \arg\min\limits_{\mathclap{\{V_j\in U| o_j<o_i\}}}\{o_i-o_j\}$;
\IF{$\textup{1}_{\rm rc}\cdot \textup{1}_{\rm aa}==1$}
\STATE $U\leftarrow U\cup \{V_i\}$ \qquad\qquad\qquad\qquad\quad\COMMENT{\% accept $V_i$}\\
\STATE $P(V_i)=\textsc{PTF}(P_{\rm left},V_{\rm left},V_i)$;
\ENDIF
\ENDWHILE
\\
\COMMENT{\% the following loop shows the \textit{ride-taking procedure}}\\
\WHILE{$U\neq \varnothing$}
\STATE The UAV flies to $o_i$ of each $V_i\in U$ right at time $t_i$;
\ENDWHILE
\IF{condition (\ref{powerenoughconditionalg}) is satisfied}
\STATE Ride-accepting procedure stops, UAV flies towards $A$ directly after taking rides in $U$.
\ENDIF
\end{algorithmic}
\end{algorithm}

\end{document}